\documentclass[a4paper,12pt]{article}

\usepackage{ifpdf}
\newif\ifpdf
\ifx\pdfoutput\undefined
  \pdffalse
\else
  \pdfoutput=1
  \pdftrue
\fi

\RequirePackage{xspace} %
\RequirePackage{subfigure} %
\RequirePackage[centertags]{amsmath} %
\RequirePackage{amssymb}
\RequirePackage{wrapfig} %
\RequirePackage{calc} %
\RequirePackage{ifthen}
\RequirePackage{tabularx} %
\RequirePackage{flafter} %
\RequirePackage{fancyhdr} %

\ifpdf
  \RequirePackage[pdftex]{color}%
  \RequirePackage{colortbl}%
  \RequirePackage{array}%
  \RequirePackage[pdftex]{graphicx}
  \RequirePackage[ pdftex, plainpages = false, pdfpagelabels,
                 pdfpagelayout = useoutlines,
                 bookmarks,
                 breaklinks = true,
                 linktocpage,
                 pagebackref,                      
                 colorlinks = true,
                 linkcolor = blue,
                 urlcolor  = blue,
                 citecolor = blue,
                 anchorcolor = blue,
                 hyperindex = true,
                 hyperfigures
                 ]{hyperref}
\else
  \RequirePackage{color}
  \RequirePackage{colortbl}
   \RequirePackage{array}
  \RequirePackage[dvips]{graphicx}
  \RequirePackage{hyperref}
  \usepackage{rotating}
\fi

\usepackage{makeidx} 
\usepackage{setspace} 
\usepackage{rotating} 
\usepackage{ecltree}
\usepackage{epic}
\usepackage{supertabular}  
\usepackage{color}
\usepackage{exscale}
\usepackage{fontenc}
\usepackage{ifthen}
\usepackage{latexsym}
\usepackage{makeidx}
\usepackage{syntonly}
\usepackage{inputenc}
\usepackage{graphicx}
\usepackage{setspace}
\usepackage{caption2}
\usepackage[english]{babel}
\usepackage[square, comma,numbers,sort&compress]{natbib}
\usepackage{hypernat}
\usepackage{boxedminipage}
\usepackage{framed}
\usepackage{longtable}
\usepackage{setspace}
\usepackage{caption2}
\usepackage{mflogo}
\usepackage{bbding}
\usepackage[all]{hypcap}

\newcommand{\note}[1]{\marginpar[left]{\singlespace \tiny #1}}
\renewcommand{\sectionmark}[1]%
      {\markright{\thesection\ #1}} 

\renewcommand{\note}[1]{}

\newcommand{\ESRFl}{European Synchrotron Radiation Facility \index{European Synchrotron Radiation Facility (ESRF)}}

\newcommand{\HEXITECl}{High Energy X-Ray Imaging Technology\index{High Energy X-Ray Imaging Technology (HEXITEC)}}
\newcommand{\HEXITECs}{HEXITEC\index{High Energy X-Ray Imaging Technology (HEXITEC)}}

\doublespace 

\begin{document}


\title{\vspace* {2.0cm} Computational Techniques for Efficient Conversion of Image Files from Area Detectors \vspace{4.0cm} }

\author{Taha Sochi\footnote{University College London, Department of Physics \& Astronomy, Gower Street, London, WC1E 6BT. Email:
t.sochi@ucl.ac.uk.} \vspace{6cm}}


\maketitle

\thispagestyle{empty} %

\newpage
\phantomsection \addcontentsline{toc}{section}{Contents} %
\tableofcontents

\newpage
\phantomsection \addcontentsline{toc}{section}{List of Figures} %
\listoffigures

\pagestyle{headings} %
\addtolength{\headheight}{+1.6pt}
\lhead[{Chapter \thechapter \thepage}]%
{{\bfseries\rightmark}}
\rhead[{\bfseries\leftmark}]%
{{\bfseries\thepage}} 
\headsep = 1.0cm               %

\newpage
\section{Abstract} \label{Abstract}

Area detectors are used in many scientific and technological applications such as particle and
radiation physics. Thanks to the recent technological developments, the radiation sources are
becoming increasingly brighter and the detectors become faster and more efficient. The result is a
sharp increase in the size of data collected in a typical experiment. This situation imposes a
bottleneck on data processing capabilities, and could pose a real challenge to scientific research
in certain areas. This article proposes a number of simple techniques to facilitate rapid and
efficient extraction of data obtained from these detectors. These techniques are successfully
implemented and tested in a computer program to deal with the extraction of X-ray diffraction
patterns from EDF image files obtained from CCD detectors.

Keywords: area detector; computational techniques; image processing; data extraction algorithms;
CCD; EDF.

\newpage
\section{Introduction}

In the recent years, the technology of detectors and data acquisition systems has witnessed a huge
revolution. This, accompanied by the wide availability of intense radiation sources such as
synchrotron radiation beams, contributed to the huge increase in the volume of data obtained in a
typical experiment. It is not unusual these days to collect hundreds of thousands of experimental
data files occupying several tera bytes of magnetic storage from a number of correlated
measurements within just a few days. This situation necessitates the development of new
computational algorithms and strategies to process and analyze such massive data sets. The current
article presents a number of simple techniques that were developed and used recently by the author
to deal with the processing of huge quantities of EDF (which stands for European Data Format)
binary image files obtained from charge-coupled devices (CCD) on synchrotron X-ray beamlines to
extract numeric data in the form of 1D diffraction patterns. These techniques are simple and
general and hence can be easily implemented and used by the interested scientists as a substitute
for commercial and free software that rely on more sophisticated but slower algorithms. In the
following we describe these techniques in the context of data extraction from binary image files of
EDF format obtained from charge-coupled devices, although they can be equally applied to other data
formats obtained by other types of detector.

\section{EDF Image Processing}

\begin{figure} [!b]
\centering
\includegraphics
[scale=0.35] {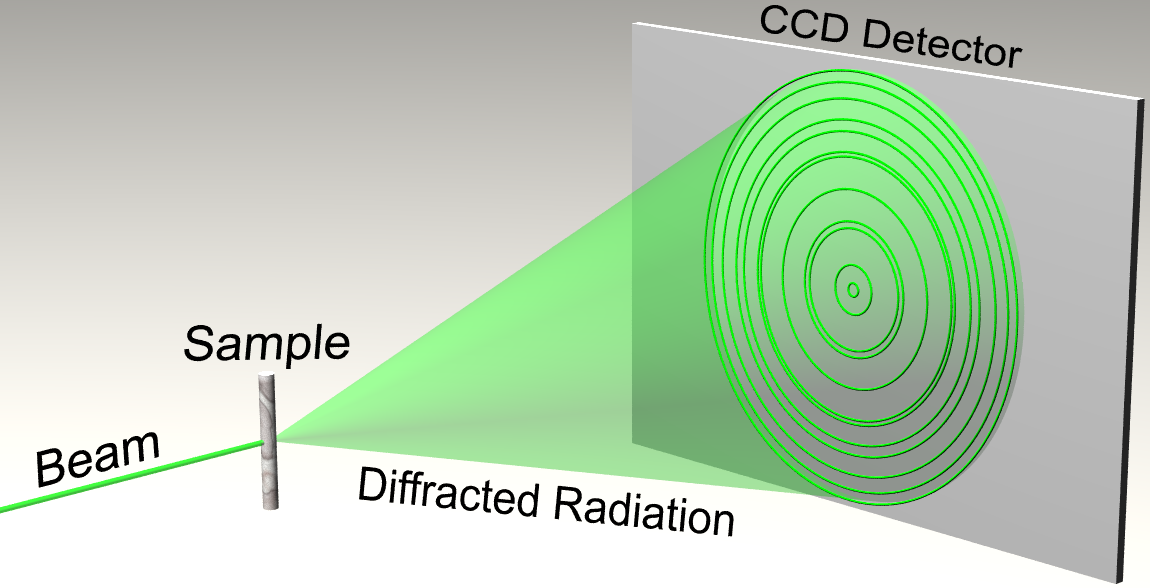}
\caption{The setting of a charge-coupled device in an X-ray diffraction experiment.}
\label{CCDSetting}
\end{figure}

Charge-coupled devices consist of an array of light-sensitive solid-state cells that convert
photons into quantified electric charges. These charges measure the intensity of the photon source
in terms of energy and number of counts. The 2D spatial distribution of these cells produce a 2D
image of the source object. Hence, an image of an object detected by a CCD device consists of a 2D
matrix of the same dimensions as the CCD array where each entry in the matrix indicates the
intensity of radiation at the corresponding cell of the CCD array. Charge-coupled devices are used
in many scientific and technological applications such as astronomy and X-ray imaging. Because CCDs
are efficient area detectors, they can reduce the acquisition time substantially with improved
resolution. A typical CCD used for X-ray imaging on a synchrotron beamline consists of an array of
more than five million cells ($2640\times1920$). Figure \ref{CCDSetting} is a simple demonstration
of the setting of a charge-coupled device in an X-ray diffraction experiment.

\begin{figure}[!h]
  \centering{}
  \includegraphics
  [width=0.9\textwidth]
  {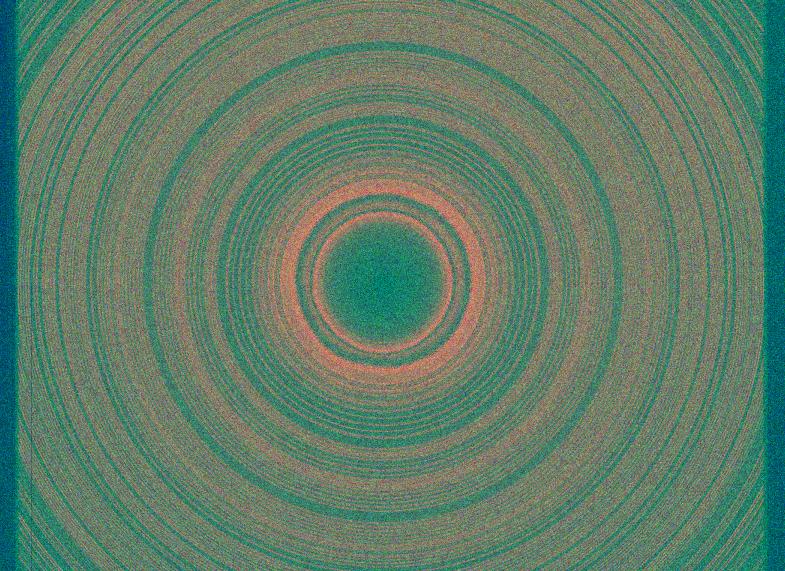}
\caption{Image of a 2D diffraction pattern contained in a typical EDF binary file. The 1D pattern
of this image is shown in Figure \ref{disCon2}.}
  \label{SampleEDF}
\end{figure}

The purpose of EDF image processing is to convert binary image files obtained from CCD detectors to
ASCII numeric format. While the binary image of an EDF file consists of a 2D rectangular matrix
where the intensity of the diffracted radiation at each cell is given as a function of implicit
$xy$ coordinates of the pixel in the matrix, the extracted ASCII numeric data represent a 1D
diffraction pattern of total intensity as a function of scattering angle. The radial dependence of
the concentric rings in the 2D pattern is correlated to the scattering angle in the 1D pattern by a
simple geometric relation. An image of the data contained in a typical EDF file is displayed in
Figure \ref{SampleEDF} while a sample of the extracted 1D pattern is shown in Figure \ref{disCon2}.
Each EDF file contains, beside the binary data of the rectangular matrix, an ASCII header which
normally consists of 24 lines of text. This header includes, among other things, the endian bit
type (little or big), the data type (e.g. unsigned short or long integer), the horizontal and
vertical dimensions of the image matrix in pixels, the size of the data file in bytes, and the date
and time of data acquisition. In the following sections we outline the main steps of data
extraction of these files.

\subsection{Radial Data Vector}

To convert the rings of pixel values in the rectangular data matrix to a 1D pattern, a radial 1D
vector is used to store the cumulative intensity of each ring. Each cell in the rectangular data
grid is assigned a certain radius according to its distance from the image center taking into
account the tilt in the horizontal and vertical orientations as will be discussed in section
\ref{TiltCorrection}. As the pixel values are read, they are assigned to the radial cells in the 1D
vector immediately. This ensures rapid processing with minimal computing resources since no memory
space is required to store the data in an intermediate processing stage. The radial dependence is
computed from the pixel implicit coordinates in the rectangular matrix when the routine is run in
single mode, while it is obtained in a more efficient way by using lookup table when the routine is
executed in a multi batch mode as will be outlined later.

Because the pixel can be too coarse as a unit of length and as a unit of intensity storage due to
its finite measurable size resulting in uneven distribution of intensity in the data points of the
1D pattern, the pixels of the CCD grid can be split to take into account these two factors. This
splitting takes two forms:

\begin{enumerate}

\item
By taking the radial unit length as a fraction of a pixel so that more than one ring can fit within
one unit pixel of radial distance. The radial assignment of a pixel then depends on its radial
distance from the image center as a float quantity rather than as an integer quantity.

\item
By splitting the intensity of the pixel according to the area contained within the closest ring so
each ring represents a strip with finite width in the radial direction. To do this, each pixel is
divided into a grid of small squares (e.g. 10$\times$10). The distance between the center of each
small square in this grid and the center of the image is then computed and allocated to the nearest
ring. The number of allocated points to a particular ring is then divided by the total number of
points in the pixel (i.e. 100 for a 10$\times$10 grid) and the fraction of the total intensity of
that pixel is added to the corresponding ring.

\end{enumerate}

\subsection{Tilt Correction} \label{TiltCorrection}

Because the CCD device may not be perpendicular to the primary beam line direction, the diffracted
ray can be deflected to a higher or lower radial distance resulting in errors in the scattering
angle. Therefore, the tilt in the vertical and horizontal orientations requires correction to
obtain accurate diffraction patterns. The derivation of this correction is presented in section
\ref{TiltCorrection}. To ensure rapid data processing and minimal computing resources, this
correction is computed only once from a representative image in any single run regardless of the
number of files processed in that run.

\subsection{Missing-Rings Correction}

Because charge-coupled devices have a rectangular shape, the rings whose radius exceeds a certain
limit, by having a radius greater than the distance between the ring center and one or more of the
rectangle sides, will not be complete (refer to the rings near the corners in Figure
\ref{SampleEDF}) and therefore the pattern will have reduced intensity at high scattering angles.
To compensate for the loss of intensity for the incomplete diffraction rings, a simple and time
saving technique is used. This technique scales the incomplete rings to the continuum value at that
radius by multiplying the total intensity of incomplete rings by the ratio of the continuum
circumference at that radius to the actual number of pixels in that ring. This number is obtained
once in any single operation while serially reading the pixel values. These scale ratios are stored
in a 1D double vector to be used at the end of each application of the extraction routine on an
individual binary data file when processing multiple files.

\begin{figure} [!h]
\centering
\includegraphics
[width=0.9\textwidth] {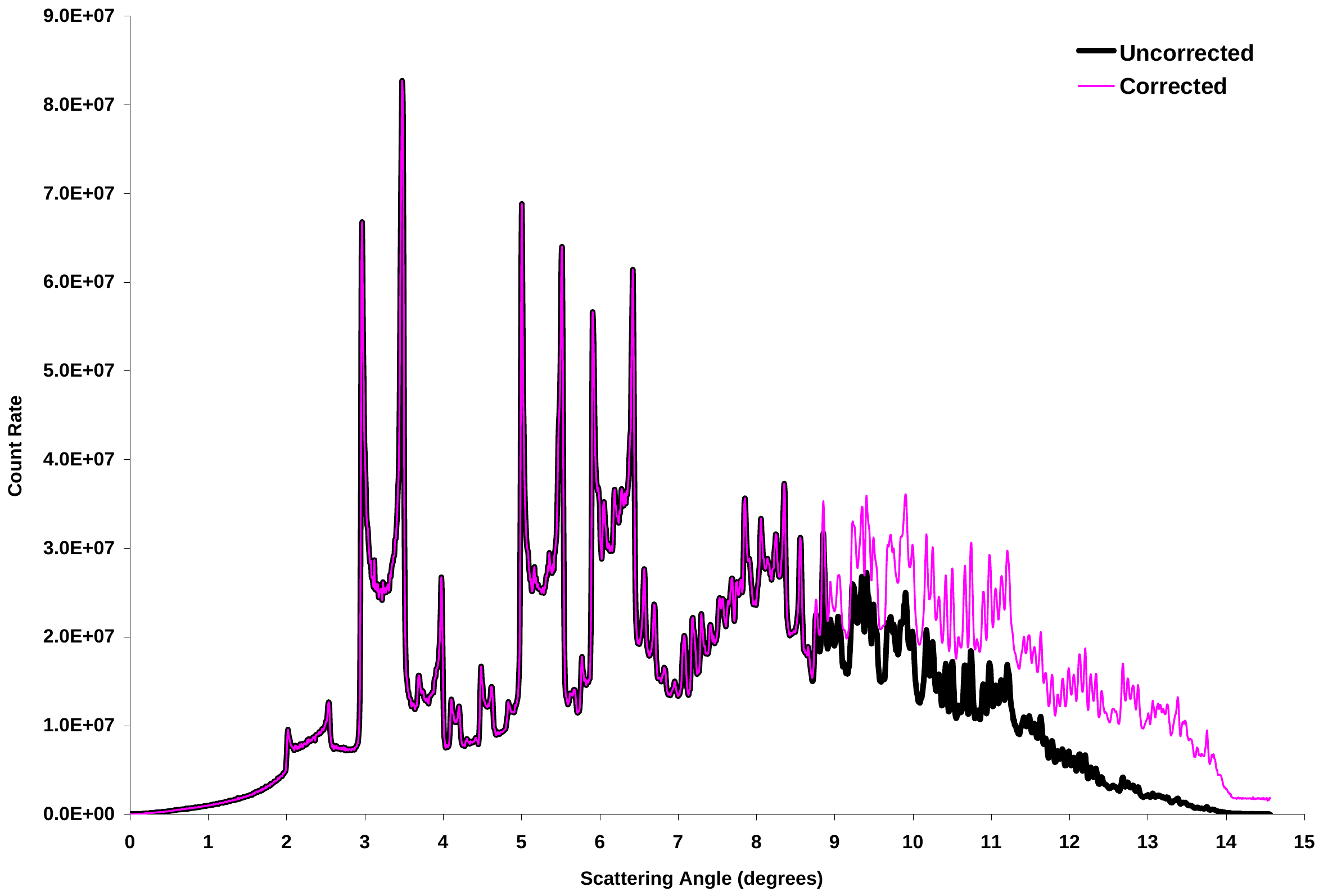}
\caption{Example of a 1D diffraction pattern extracted from a 2D EDF pattern. The pink curve is
obtained with the application of missing-rings correction while the black is obtained without this
correction.} \label{disCon2}
\end{figure}

\begin{figure} [!h]
\centering
\includegraphics
[width=0.9\textwidth] {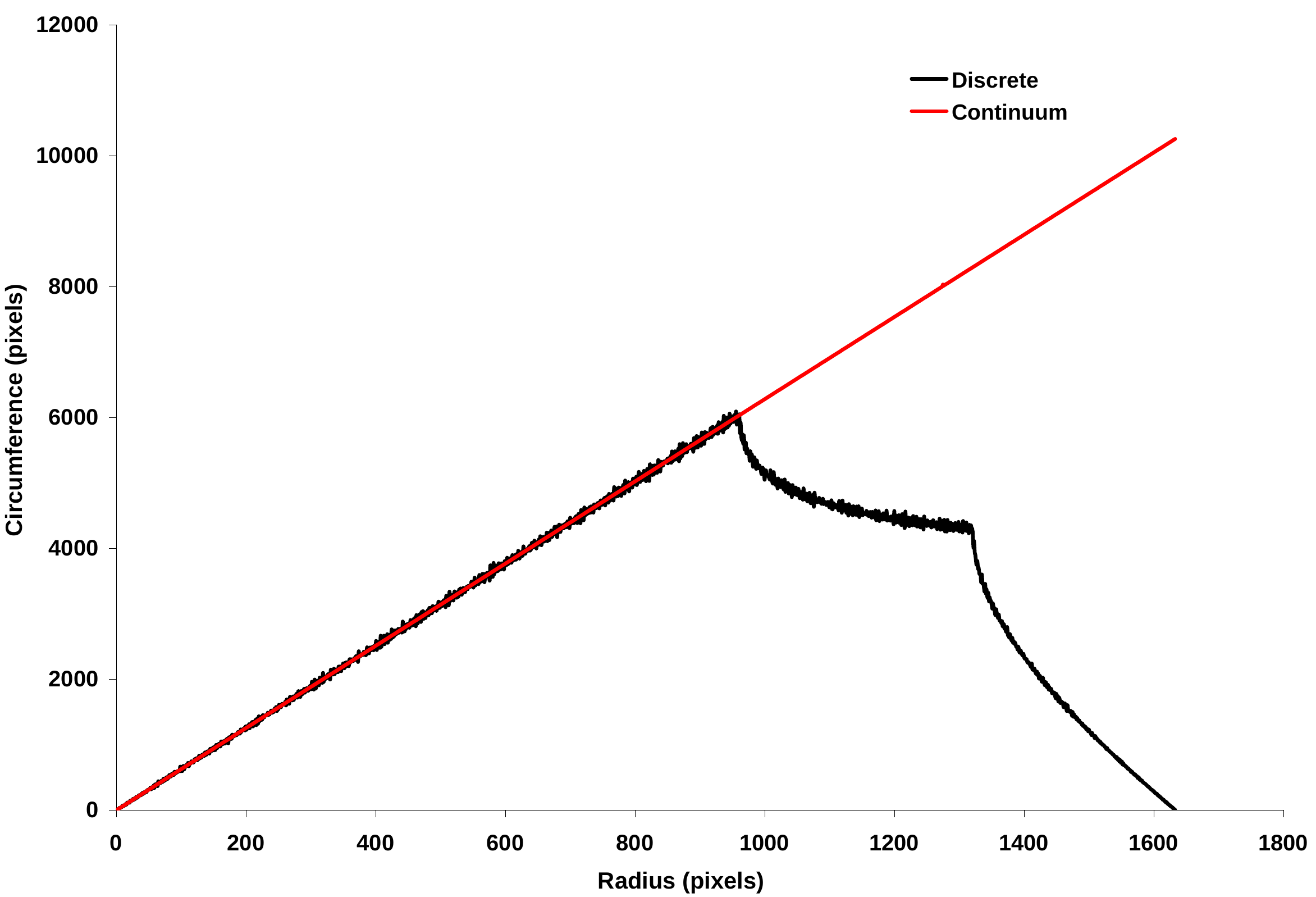}
\caption{The continuum approximation alongside the discrete value for the number of pixels in a
ring.} \label{disCon}
\end{figure}

Figure \ref{disCon} displays an example of the actual (discrete) number of pixels of the image
rings as a function of radius in pixels alongside the continuum value of $2\pi r$ represented by
the straight line. As can be seen, the two curves match very well for the complete rings. This
indicates that the continuum is a very good approximation apart from the rings that are too close
to the image corners. The meaning of the two cusps is obvious as the radius of the rings increases
and exceeds the rectangle sides in one direction (i.e. vertical or horizontal) and then in the
other direction. Figure \ref{disCon2} presents a sample diffraction pattern obtained from an EDF
image with and without the application of missing-rings correction.

\subsection{Multi Batch Operation}

One of the strategies that we employ in EDF image extraction routine is multi batch processing
where the routine is applied in a single run on a data directory that includes multiple data sets
each stored within a specific folder. This multi batch processing is applied in an organized
fashion and the resulted data are archived systematically using meaningful naming to facilitate
managing and post precessing. As a time-saving measure, when EDF data extraction routine is run in
a multi batch mode the header is read only once from a representative image file to obtain the
required information. Another time saving measure used in multi batch operations is the use of
lookup matrix for radial assignment of the individual pixels. These assignments are computed only
once at the start of operation and stored in a 2D vector for the use in the subsequent data
extraction operation without repeating these lengthy calculations. The radial assignment of pixels
include tilt correction in the horizontal and vertical directions. In multi batch mode the scale
ratios which are required for missing-rings correction are also computed once at the start of
operation and stored in a 1D lookup vector as indicated already. All these measures ensure rapid
data processing and considerable save in computational resources.

\section{Case Study} \label{CaseStudy}

The data extraction techniques which we described have been implemented in a rapid-analysis
software called EasyDD \cite{scienceware}. EasyDD has been used in a number of studies (e.g.
\cite{LazzariJSB2009, EspinosaOJBJe2009}), and is in use by the \HEXITECl\ (\HEXITECs) project
\cite{HEXITEC}. The EDF extraction algorithm requires as an input the CCD image center, the
detector tilt and the physical dimensions of the detector system. The user has the opportunity to
select the type of the extraction operation by determining the radial size of the extracted
pattern, the application of missing-rings correction and the radial split factor.

\begin{figure} [!h]
\centering
\includegraphics
[scale=0.9] {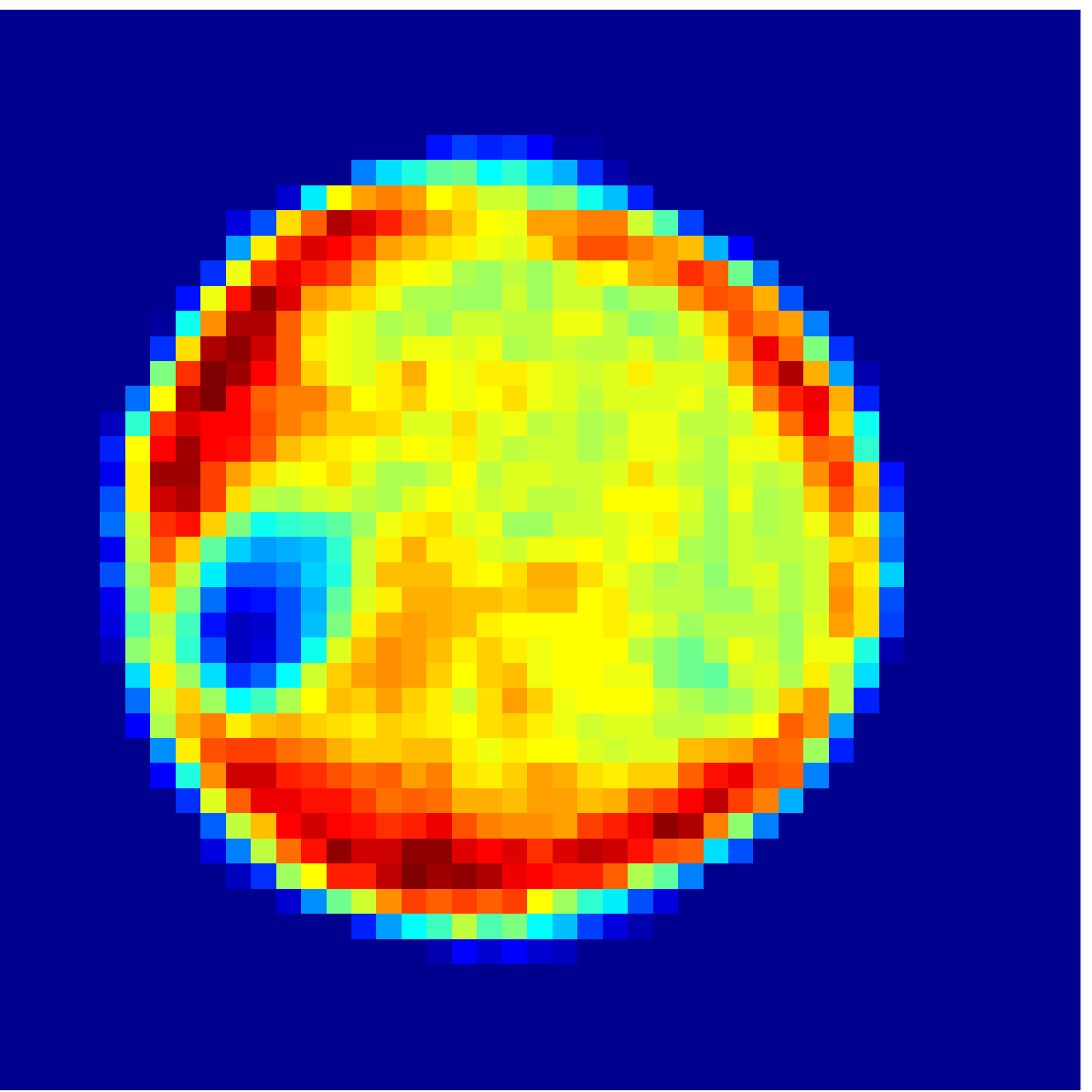}
\caption{A tomographic image of a diffraction peak obtained from a nickel compound on a cylindrical
alumina extrudate sample. The numeric data are obtained from EasyDD using EDF extraction, back
projection and curve-fitting routines.} \label{Example}
\end{figure}

In a recent study, EasyDD has been used for extracting and processing data obtained from the
\ESRFl\ (ESRF). The measurements were carried out at the beamline ID15B which is dedicated to
applications that require very high energy X-ray radiation up to several hundreds of keV
\cite{ESRF}. The data, which were collected over a few days, consist of about 254 thousand EDF
image files in 179 data sets with total size of about 2.45 tera bytes. EasyDD was used to extract
the data and convert these images to 1D spectral patterns in ASCII numeric format. This was
performed, using multi-batch mode, in about 36 hours of CPU time on an ordinary desktop computer.
This time, when compared to an estimated 2 months on a rival commercial software, reveals the
efficiency of our EDF extraction techniques and the crucial role that they can play in rapid
processing of huge data sets. A sample of these data sets is presented in Figure \ref{Example} for
one of the diffraction peaks.

Finally, it should be remarked that despite the fact that the data conversion algorithm in its
current state is for numeric conversion of EDF files, it can be easily extended to other data types
with similar structure. The algorithm can also convert the EDF binary images to 2D visual images in
a number of formats (png, jpg and bmp). An example of these images is given in Figure
\ref{SampleEDF}. This type of operation is approximately as fast as numeric conversion to 1D
patterns. The algorithm can also perform fidelity check to find and compensate for the missing
files in an integrated data set consisting of files with regular naming pattern.

\clearpage
\renewcommand{\refname}{}

\newpage
\section{Appendix: Tilt Correction Derivation} \label{TiltCorrection}

In this derivation we use a standard Cartesian coordinate system in a plane perpendicular to the
original beam direction that passes through the center of the image and hence the origin of the
coordinate system. For simplicity we use a unit circle centered at the origin and lying in this
plane. In the following derivation we have:

$\phi$ is the angle of an arbitrary point on the unit circle ($0 \leq \phi < 2\pi$)

$\theta_x$ is the tilt in the $x$ direction

$\theta_y$ is the tilt in the $y$ direction

$\theta$ is the actual tilt at an arbitrary point on the tilted CCD image. It is obvious that
$\theta$ is a function of $\phi$ and can have negative as well as positive values.

From a simple geometric argument, the actual tilt as a function of $\phi$ is given by
\begin{equation}
    \theta(\phi) = \arctan[\cos \phi \times \tan \theta_x + \sin \phi \times \tan \theta_y]
\end{equation}

As can be seen, this formula retrieves the correct (and obvious) tilt in the $x$, $y$, $-x$ and
$-y$ directions.

\begin{figure}[!h]
  \centering{}
  \includegraphics
  [width=0.55\textwidth]
  {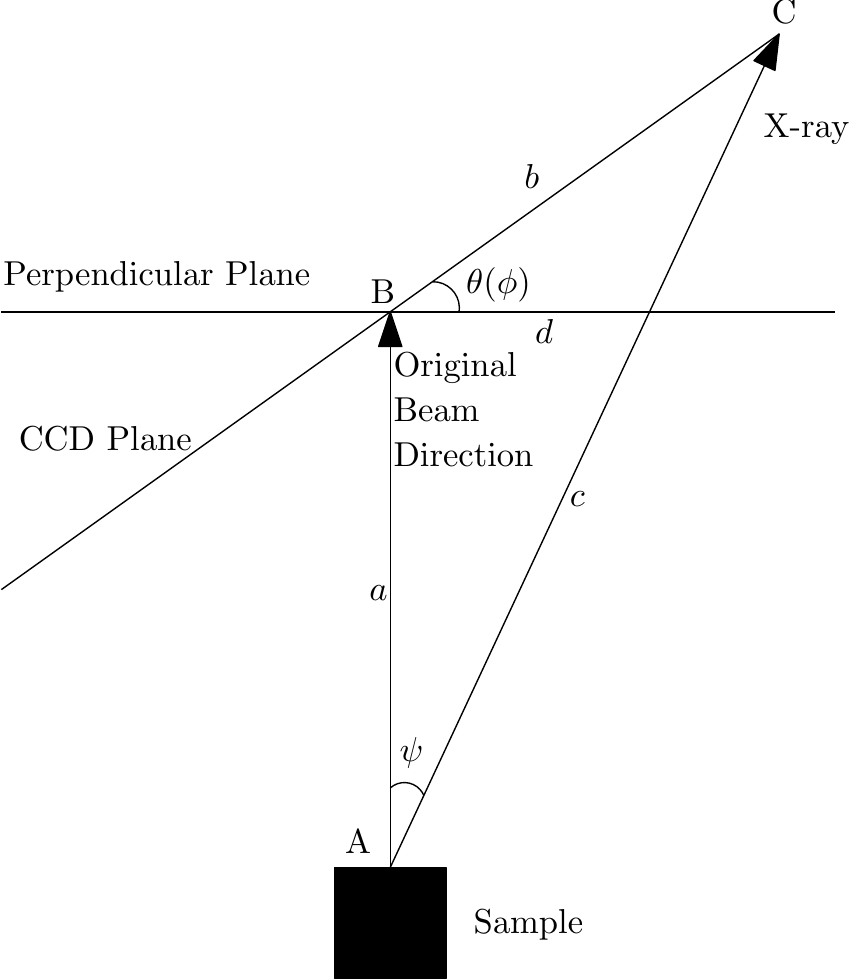}
  \caption{A schematic diagram for demonstrating the derivation of tilt correction.}
  \label{tiltCor}
\end{figure}

Now referring to Figure \ref{tiltCor}, which is in the plane identified by the original beam and
ray directions, and on applying the cosine rule on the triangle ABC we have

\begin{equation}
    c = \sqrt{a^2 + b^2 - 2ab \cos \left(\theta + \frac{\pi}{2} \right)}
\end{equation}

Using the sine rule we obtain the scattering angle
\begin{equation}
    \psi = \arcsin \left[b \times \frac{\sin (\theta+\pi/2)}{c} \right]
\end{equation}

And finally the required quantity, $d$, can be obtained from
\begin{equation}
    d = a \times \tan \psi
\end{equation}

It is obvious that there are other methods for finding $d$. Although this derivation is
demonstrated for $\theta > 0$, as seen in Figure \ref{tiltCor}, it is general and valid for
$-\frac{\pi}{2} < \theta < \frac{\pi}{2}$.

\end{document}